# Metamaterial model of fractal time

Igor I. Smolyaninov

*Department of Electrical and Computer Engineering, University of Maryland, College Park, MD 20742, USA*

*phone: 301-405-3255; fax: 301-314-9281; e-mail: smoly@umd.edu*

**While numerous examples of fractal spaces may be found in various fields of science, the flow of time is typically assumed to be one-dimensional and smooth. Here we present a metamaterial-based physical system, which can be described by effective three-dimensional (2+1) Minkowski spacetime. The peculiar feature of this system is that its time-like variable has fractal character. The fractal dimension of the time-like variable appears to be D=2.**

Recent advances in metamaterial optics [1-3] enable researchers to design novel physical systems, which can be described by effective space-times having very unusual metric and topological properties. Examples include constructing analogues of black holes [4-7], wormholes [8,9], spinning cosmic strings [10], and even the metric of Big Bang itself [11]. More unusual examples include building a physical system, which can be described by (-,-,+,+) metric signature, which differs from the usual Minkowski signature (-,+,+,+) of physical spacetime [12]. The latter advance is based on the unusual optics of hyperbolic metamaterials, which have different signs of dielectric permittivity $\varepsilon$ along different spatial directions. Mapping of monochromatic extraordinary light distribution in a hyperbolic metamaterial along some spatial direction may model the "flow of time" in a three dimensional (2+1) effective Minkowski spacetime [11]. To better understand this effect, let us start with a non-magnetic uniaxial anisotropic material with dielectric permittivities $\varepsilon_x= \varepsilon_y= \varepsilon_1$ and $\varepsilon_z = \varepsilon_2$, and assume that this behaviour holds in some frequency range around $\omega=\omega_0$. Any electromagnetic field propagating in this material can be expressed as a sum of the "ordinary" and "extraordinary" contributions, each of these being a sum of an arbitrary number of plane waves polarized in the "ordinary" ($\vec{E}$ perpendicular to the optical axis) and "extraordinary" ($\vec{E}$ parallel to the plane defined by the k–vector of the wave and the optical axis) directions. Let us define our "scalar" extraordinary wave function as $\varphi=E_z$ so that the ordinary portion of the electromagnetic field does not contribute to $\varphi$. Since metamaterials exhibit high dispersion, let us work in the frequency domain and write the macroscopic Maxwell equations as [13]

$$\frac{\omega^2}{c^2}\vec{D}_\omega = \vec{\nabla}\times\vec{\nabla}\times\vec{E}_\omega \text{ and } \vec{D}_\omega = \vec{\varepsilon}_\omega \vec{E}_\omega, \tag{1}$$

which results in the following wave equation for $\varphi_\omega$ if $\varepsilon_1$ and $\varepsilon_2$ are kept constant inside the metamaterial:

$$-\frac{\omega^2}{c^2}\varphi_\omega = \frac{\partial^2\varphi_\omega}{\varepsilon_1\partial z^2} + \frac{1}{\varepsilon_2}\left(\frac{\partial^2\varphi_\omega}{\partial x^2} + \frac{\partial^2\varphi_\omega}{\partial y^2}\right) \tag{2}$$

In hyperbolic metamaterials [14] $\varepsilon_1$ and $\varepsilon_2$ have opposite signs. Let us consider the case of constant $\varepsilon_1 >0$ and $\varepsilon_2 <0$, and assume that this behavior holds in some frequency range around $\omega=\omega_0$. Such behavior can be obtained in the "layered" hyperbolic metamaterial [15] shown in Fig.1(a). Let us assume that the metamaterial is illuminated by coherent CW laser field at frequency $\omega_0$, and we study spatial distribution of the extraordinary field $\varphi_\omega$ at this frequency. Under these assumptions equation (2) can be re-written in the form of 3D Klein-Gordon equation describing a massive scalar $\varphi_\omega$ field:

$$-\frac{\partial^2\varphi_\omega}{\varepsilon_1\partial z^2}+\frac{1}{|\varepsilon_2|}\left(\frac{\partial^2\varphi_\omega}{\partial x^2}+\frac{\partial^2\varphi_\omega}{\partial y^2}\right)=\frac{\omega_0{}^2}{c^2}\varphi_\omega=\frac{m^{*2}c^2}{\hbar^2}\varphi_\omega \qquad (3)$$

in which the spatial coordinate $z=\tau$ behaves as a "timelike" variable. Therefore, eq.(3) describes world lines of massive particles which propagate in a flat (2+1) Minkowski spacetime. When a metamaterial is built and illuminated with a coherent extraordinary CW laser beam, the stationary pattern of light propagation inside the metamaterial represents a complete "history" of a toy (2+1) dimensional spacetime populated with particles of mass $m^*$ (Fig.1(b)). This "history" is written as a collection of particle world lines along the "timelike" $z$ coordinate. The effective interval in this spacetime may be written as

$$ds^2=-\varepsilon_1 dz^2+|\varepsilon_2|\left(dx^2+dy^2\right)\ , \qquad (4)$$

where the "time interval" between events may be calculated as

$$T=\int d\tau=\int \varepsilon_1^{1/2}dz \qquad (5)$$

Now let us recall that in addition to frequency dispersion, hyperbolic metamaterials typically exhibit spatial dispersion, which means that $\varepsilon_1$ and $\varepsilon_2$ exhibit weak dependence on the wave vector of the form

$$\varepsilon=\varepsilon^{(0)}+\alpha\left(\frac{k^2c^2}{\omega^2}\right) \qquad (6)$$

where $\alpha$ is small [16]. When the absolute values of $\varepsilon^{(0)}{}_1$ and $\varepsilon^{(0)}{}_2$ are large, spatial dispersion can be disregarded, and description of the hyperbolic medium in terms of effective (2+1) Minkowski spacetime is a good approximation. On the other hand, the effect of spatial dispersion may become very interesting if either $\varepsilon_1$ or $\varepsilon_2$ are close to zero. Below we will demonstrate that the case of constant small positive $\varepsilon_1 \sim 0$ and large negative $\varepsilon_2$ leads to very interesting and peculiar physical situation, in which the behavior of time-like $z=\tau$ variable becomes fractal.

Fractal objects and fractal dimensions are very useful mathematical concepts. A typical problem where the fractal dimension arises naturally is an attempt to measure the perimeter of an island in the ocean. The result would depend on the resolution used in the measurements. The value of the perimeter measured on the large scale from an aerial map would be much smaller than the value obtained by walking along the beach with a ruler, when every tiny curve of the beach is measured. The fractal dimension $D$ is defined from the variation with resolution of the main fractal variable (a length $L$ of a fractal curve, an area of a fractal surface, etc.) [17]. If $D_T$ is the topological dimension ($D_T$=1 for a curve, $D_T$=2 for a surface, etc.), the scale or fractal dimension $D=D_T+\delta$ is defined as

$$\delta = \frac{d(\ln L)}{d\left(\ln(l/\lambda)\right)} \quad , \tag{7}$$

where $\lambda$ is the resolution of the measurements. If $\delta$ is constant and $D_T$=1 we obtain

$$L = L_0 \left(\frac{l}{\lambda}\right)^{\delta} \quad , \tag{8}$$

where the length $L_0$ is measured when $\lambda = l$. A well known example of a fractal self-similar curve (the Koch snowflake having fractal dimension $D$=1.2619) is shown in

Fig.2. Compactified fractal extra dimensions of space have been considered in [18], while theoretical models of fractal time and more general models of fractal space-time has been suggested by many authors since about 30 year ago [19-22]. These models may help tame divergencies, which typically occur in the quantum gravity theories on sub-planckian scales. However, the models of fractal spacetime have remained in the theoretical domain only.

Let us now examine the behavior of "time intervals" (eq.(5)) in a hyperbolic metamaterial in the case of constant small positive $\varepsilon_1 \sim 0$ and large negative $\varepsilon_2$ . Neglecting spatial dispersion of $\varepsilon_2$, and assuming

$$\varepsilon_1 = \varepsilon_1^{(0)} + \alpha\left(\frac{k_z^{\ 2} c^2}{\omega^2}\right) \approx \alpha\left(\frac{k_z^{\ 2} c^2}{\omega^2}\right) \qquad (9)$$

the time intervals are equal to

$$T = \int d\tau = \int \varepsilon_1^{1/2} dz = \int \alpha^{1/2} \frac{k_z c}{\omega} dz = T_0\left(\frac{2\pi c}{\omega\lambda}\right) \qquad (10)$$

which indicates that the measured time intervals are scale-dependent. According to eq.(8), $\delta$=1 and the fractal dimension of the time-like variable appears to be $D$=2. Thus, the fractal behavior of the time-like variable found here is best illustrated as a Hilbert curve (a particular example of Peano curve, a continuous space-filling curve having D=2, which is shown in Fig.3). Note that the main physical reason for spatial dispersion in hyperbolic metamaterials is the presence of surface plasmon (SP) modes propagating along metal-dielectric interfaces. At large SP momenta $k_z \sim k_{xy}$ [23]. Therefore, the chosen functional dependence of $\varepsilon_1$ on $k_z$ (eq.(9)) appears natural and reasonable.

Let us consider a practical design of such metamaterial. Diagonal components of dielectric tensor of the layered hyperbolic metamaterial shown in Fig.1(a) may be obtained using Maxwell-Garnett approximation [15]:

$$\varepsilon_1 = \alpha\varepsilon_m + (1-\alpha)\varepsilon_d \quad , \qquad \varepsilon_2 = \frac{\varepsilon_m \varepsilon_d}{(1-\alpha)\varepsilon_m + \alpha\varepsilon_d} \tag{11}$$

where $\alpha$ is the fraction of metallic phase, and $\varepsilon_m<0$ and $\varepsilon_d>0$ are the dielectric permittivities of metal and dielectric, respectively. $\varepsilon_1 \sim 0$ requirement leads to the following choice of $\alpha$:

$$\alpha = \frac{\varepsilon_d}{(\varepsilon_d - \varepsilon_m)} \quad , \tag{12}$$

which leads to $\varepsilon_2$ being equal to

$$\varepsilon_2 = \frac{\varepsilon_m \varepsilon_d (\varepsilon_d - \varepsilon_m)}{(\varepsilon_d^2 - \varepsilon_m^2)} \quad , \tag{13}$$

We can see that $\varepsilon_2<0$ requirement may be satisfied if $-\varepsilon_m<\varepsilon_d$. Therefore, such low loss “alternative plasmonic materials” [24] as indium tin oxide (ITO) can be used as the “metallic” component of the required hyperbolic metamaterial in the near IR (1.5-2 μm) spectral range.

Building an experimental model of fractal time and more general fractal space-time will provide us with an interesting tool to gain direct experimental insights into the “analogue sub-planckian physics”. Modern developments in gravitation research strongly indicate that classic general relativity is an effective macroscopic field theory, which needs to be replaced with a more fundamental theory based on yet unknown microscopic degrees of freedom. It seems plausible that the macroscopic general relativity domain, and the microscopic physics domain are separated by an interesting region in which space-time behavior may indeed show fractal features. Therefore, experimental studies of such an analogue transitional region could be extremely useful. Moreover, recent observation that physical vacuum itself may behave as a hyperbolic metamaterial [25] provides an additional incentive.

In conclusion, we have presented a “layered” hyperbolic metamaterial system exhibiting spatial dispersion, which can be described by effective three-dimensional (2+1) Minkowski spacetime. The peculiar feature of this system is that its time-like variable has fractal character. The fractal dimension of the time-like variable appears to be $D$=2. Such a metamaterial may be realized in the near IR range using such “alternative plasmonic materials” as ITO.

**References**

[1] J. B. Pendry, D. Schurig, D.R. Smith, “Controlling electromagnetic fields”, *Science* **312**, 1780-1782 (2006).

[2] U. Leonhardt, “Optical conformal mapping”, *Science* **312**, 1777-1780 (2006).

[3] U. Leonhardt and T. G. Philbin, “General Relativity in Electrical Engineering”, *New J. Phys.* **8**, 247 (2006).

[4] I.I. Smolyaninov, “Surface plasmon toy-model of a rotating black hole”, *New Journal of Physics* **5**, 147 (2003).

[5] D.A. Genov, S. Zhang, and X. Zhang, “Mimicking celestial mechanics in metamaterials”, *Nature Physics*, **5**, 687-692 (2009).

[6] E.E. Narimanov and A.V. Kildishev, “Optical black hole: Broadband omnidirectional light absorber”, *Appl. Phys. Lett.* **95**, 041106 (2009).

[7] Q. Cheng and T. J. Cui, “An electromagnetic black hole made of metamaterials”, arXiv:0910.2159v3

[8] A. Greenleaf, Y. Kurylev, M. Lassas, and G. Uhlmann, “Electromagnetic wormholes and virtual magnetic monopoles from metamaterials”, *Phys. Rev. Lett.* **99**, 183901 (2007).

[9] I.I. Smolyaninov, “Metamaterial “Multiverse””, *Journal of Optics* **13,** 024004 (2010).

[10] T. G. Mackay and A. Lakhtakia, “Towards a metamaterial simulation of a spinning cosmic string”, *Phys. Lett. A* **374**, 2305-2308 (2010).

[11] I.I. Smolyaninov and Y.J. Hung, “Modeling of time with metamaterials”, *JOSA B* **28**, 1591-1595 (2011).

[12] I.I. Smolyaninov and E.E. Narimanov, “Metric signature transitions in optical metamaterials”, *Phys. Rev. Letters* **105**, 067402 (2010).

[13] L.Landau, E.Lifshitz, Electrodynamics of Continuous Media (Elsevier, 2004).

[14] Z. Jakob, L.V. Alekseyev, and E. Narimanov, “Optical hyperlens: far-field imaging beyond the diffraction limit”, *Optics Express* **14**, 8247-8256 (2006).

[15] R. Wangberg, J. Elser, E.E. Narimanov, and V.A. Podolskiy, “Nonmagnetic nanocomposites for optical and infrared negative-refractive-index media”, *J. Opt. Soc. Am. B* **23**, 498-505 (2006).

[16] V. M. Agranovich and V. L. Ginzburg, Spatial Dispersion in Crystal Optics and the Theory of Excitons (Wiley, 1966).

[17] B. Mandelbrot, Fractals: form, chance, and dimension (Freeman, San Francisco, 1977).

[18] I.I. Smolyaninov, "Fractal extra dimension in Kaluza-Klein theory", *Phys.Rev.D* **65**, 047503 (2002).

[19] O. Lauscher and M. Reuter, “Fractal spacetime structure in asymptotically safe gravity”, *JHEP* **10**, 050 (2005).

[20] K. Svozil, “Quantum field theory on fractal spacetime: a new regularisation method”, *J. Phys. A: Math. Gen*. **20**, 3861 (1987).

[21] M. F. Shlesinger, “Fractal time in condensed matter”, *Ann. Rev. Phys. Chem.* **39**, 269-290 (1988).

[22] G. N. Ord, “Fractal space-time: a geometric analogue of relativistic quantum mechanics”, *J. Phys. A: Math. Gen.* **16** 1869 (1983).

[23] A.V. Zayats, I.I. Smolyaninov, and A. Maradudin, "Nano-optics of surface plasmon-polaritons", *Physics Reports* **408**, 131-314 (2005).

[24] G.V. Naik, J. Kim, and A. Boltasseva, “Oxides and nitrides as alternative plasmonic materials in the optical range”, *Opt. Mater. Express* **1**, 1090-1099 (2011).

[25] I.I. Smolyaninov, “Vacuum in strong magnetic field as a hyperbolic metamaterial”, *Phys. Rev. Letters* **107**, 253903 (2011) .

**Figure Captions**

**Figure 1.** (a) Schematic view of the "layered" hyperbolic metamaterial made of subwavelength metal and dielectric layers. (b) Schematic view of a particle world line in a (2+1) dimensional Minkowski spacetime.

**Figure 2.** Example of a self-similar fractal curve (Koch snowflake). The fractal dimension of this curve is $D$=1.2619. Similar behavior of the time-like coordinate with $D$=2 may be achieved in a layered hyperbolic metamaterial based on ITO.

**Figure 3.** Fractal behaviour of the time-like $z$ coordinate found in the metamaterial example considered here is best illustrated as a Hilbert curve (a particular example of Peano curve, a continuous space-filling curve having D=2). First, second and third order Hilbert curves are shown as red, blue and black curves, respectively.

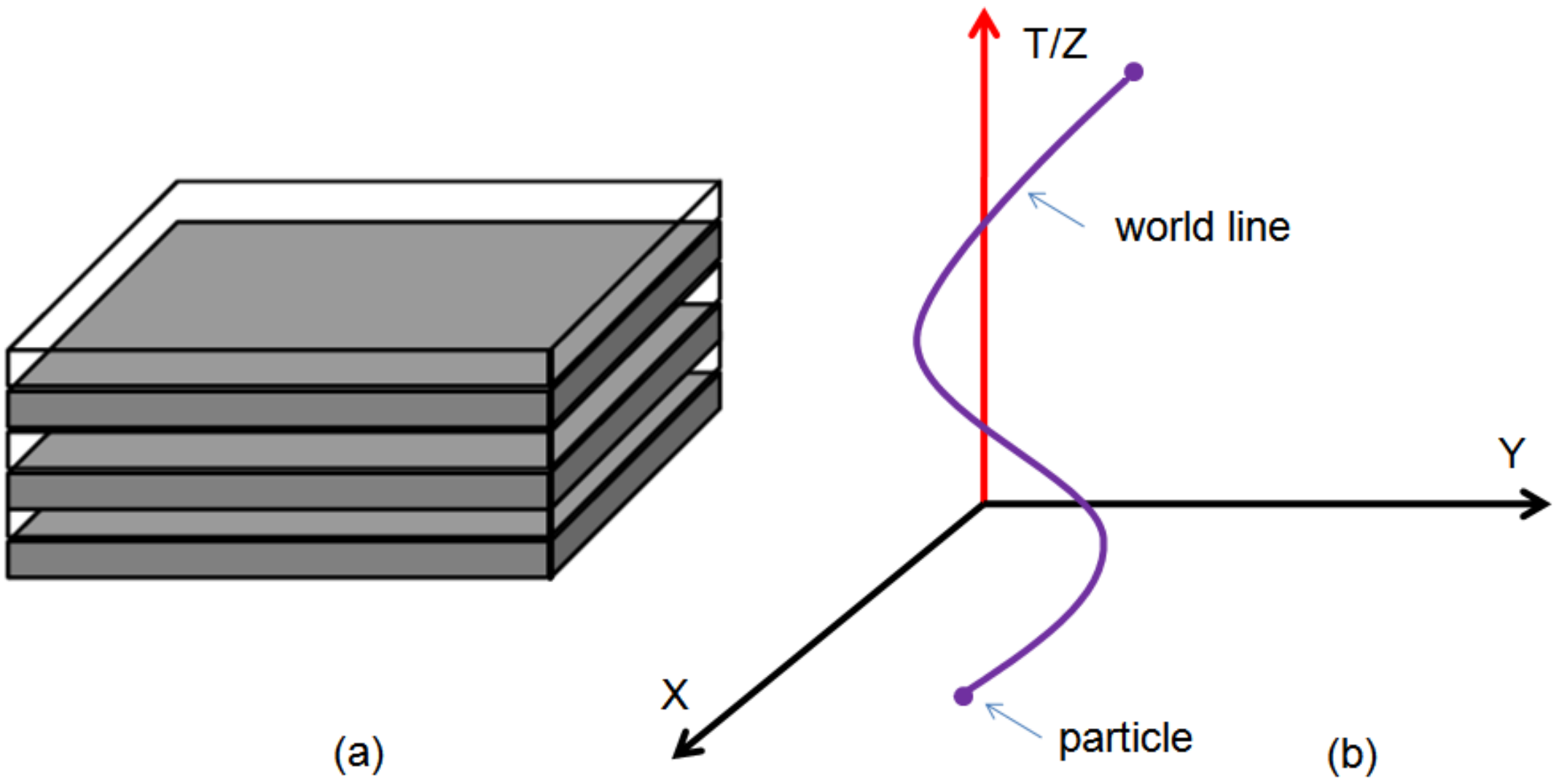

Fig.1

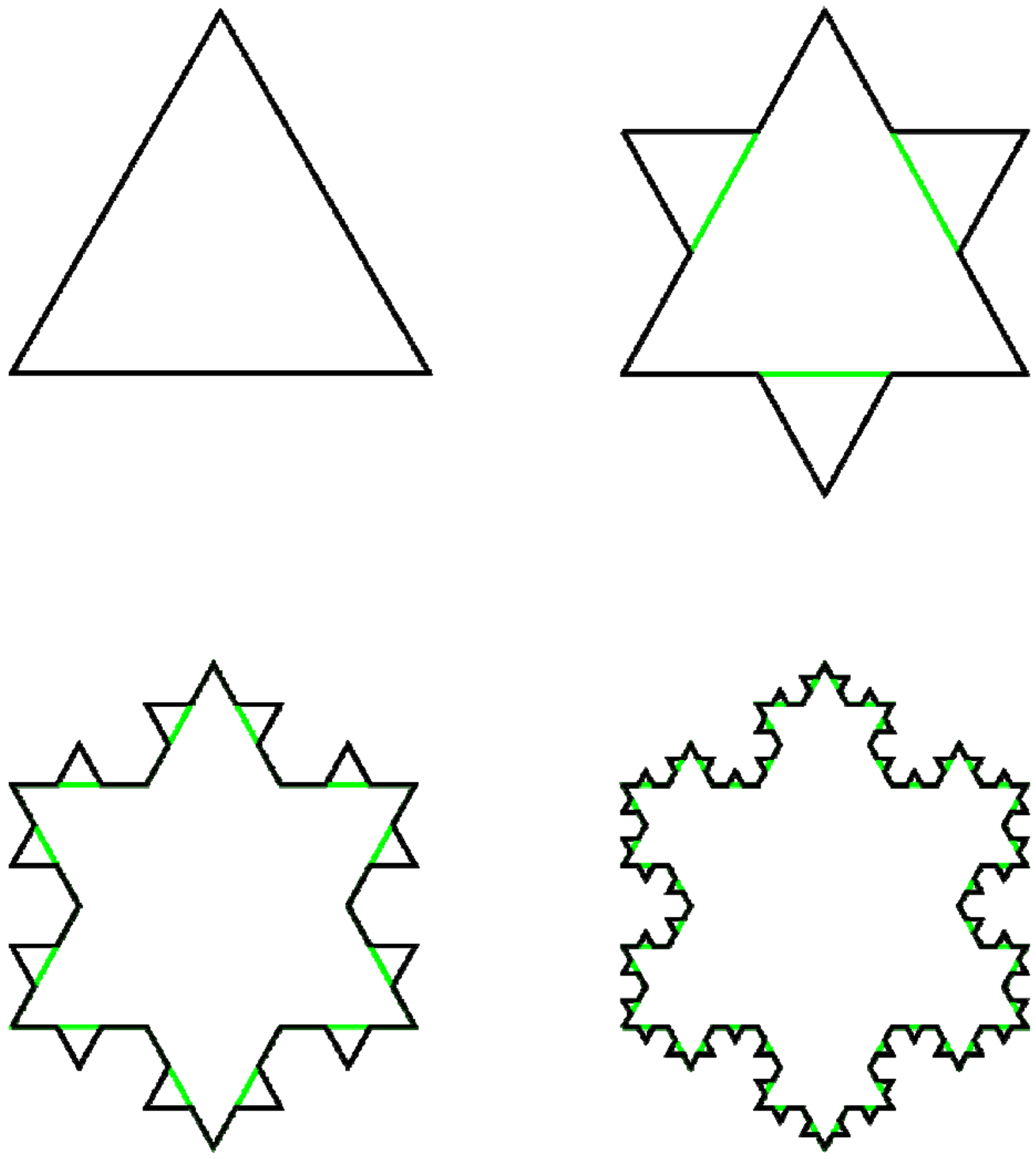

Fig.2

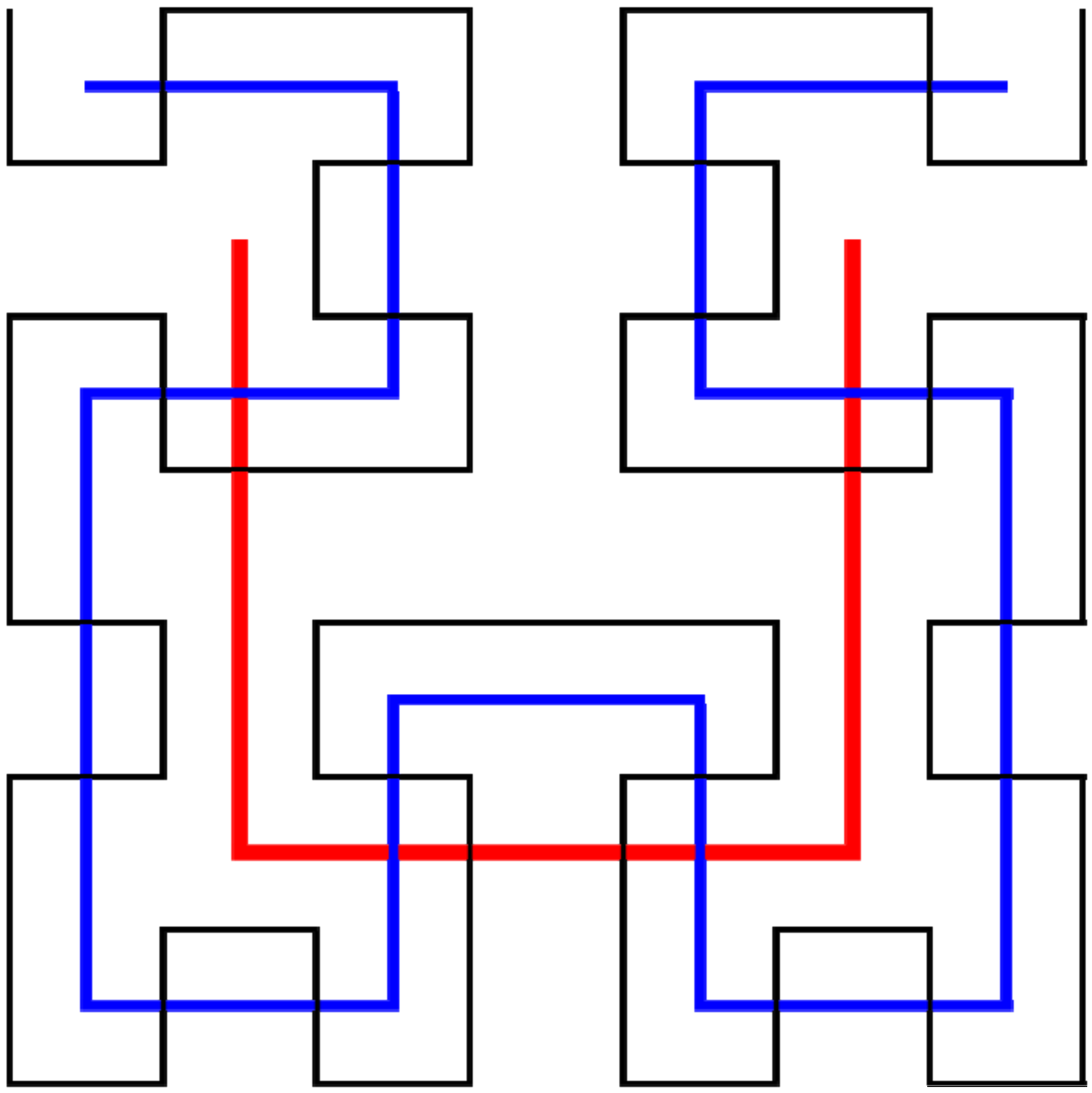

Fig.3